\documentclass[aps,prl,showpacs,twocolumn]{revtex4}

\usepackage{amssymb}
\usepackage{graphicx}
\usepackage{epsfig}
\begin{document}

\date{\today}

\title{The Statistics of the Work Done on a Quantum Critical System by Quenching a Control Parameter.}
\author{ Alessandro Silva}
\affiliation{Abdus Salam ICTP, Strada Costiera 11, 34100 Trieste,
Italy} \pacs{05.70.Ln, 03.65.Yz.}

\begin{abstract}
We study the statistics of the work done on a quantum critical
system by quenching a control parameter in the Hamiltonian. We
elucidate the relation between the probability distribution of the
work and the Loschmidt echo, a quantity emerging usually in the
context of dephasing. Using this connection we characterize the
statistics of the work done on a quantum Ising chain by quenching
locally or globally the transverse field. We show that for local
quenches starting at criticality the probability distribution of the
work displays an interesting edge singularity.
\end{abstract}

\maketitle

A series of recent experiments with cold atomic
gases~\cite{Greiner,Kinoshita} spurred new interest on the dynamics
of quantum correlated systems. A number of fundamental issues on the
nonequilibrium physics of quantum systems are being addressed,
ranging from the relation between nonintegrability and
thermalization~\cite{Rigol}, to the universality of defect
production for adiabatic quenches across quantum critical
points~\cite{Zurek}. In this broad context, a paradigmatic example
of experimental protocol is the instantaneous quench: an abrupt
change, either global or local, of a control parameter $g$ from some
initial value $g_0$ to a final one $g_1$. Experimentally, it has
been shown that the dynamics after such quenches may show intriguing
features, such as collapse and revivals of the order parameter for
quenches done through a quantum critical
point~\cite{Greiner,Altman}, as well as the absence of
thermalization in systems close to
integrability~\cite{Kinoshita,Rigol}.

Theoretically, the study of quantum quenches received considerable
interest: after a series of classic works on the nonequilibrium
dynamics of the quantum Ising model~\cite{Mazur}, recent
investigations focused on characterizing the long time asymptotics
of correlation functions~\cite{ Sachdev1, Igloi}, their behavior as
compared to their thermal counterparts~\cite{Rigol}, and the
universality emerging in the quench dynamics at a quantum critical
point~\cite{Cardy}. Partial information on the internal dynamics of
the system can be obtained in a variety of ways. One may extract the
way in which excitations propagate by looking at the time dependence
of correlators after a quench~\cite{Igloi,Cardy}. More subtle
information on the establishment of quantum correlations can be
obtained by studying the dynamics of entanglement
entropies~\cite{Calabrese}. The purpose of this Letter is to discuss
a more basic way to characterize both the internal dynamics and the
quench protocol itself by obtaining information on how far from
equilibrium the system has been taken. This can be done by studying
the statistics of a fundamental quantity: the work $W$ done on the
system by changing its parameters.

The main observation behind this proposal is that the quench
protocol resembles a standard thermodynamic transformation. However,
since a quench takes the system out of equilibrium, the work $W$,
unlike in a quasistatic process, is characterized by a probability
distribution $P(W)$~\cite{Jarzynski,Kurchan,Talkner}. Below, we
focus on the characteristic function of $P(W)$, defined as
\begin{eqnarray}
G(t)=\int dW e^{i W t} P(W),
\end{eqnarray}
and study it for the prototypical example of quantum critical
system, the quantum Ising chain. We first elucidate a useful
relation between $G(t)$ and the Loschmidt echo, a quantity emerging
in various physical contexts, most notably the Fermi edge
singularity~\cite{Schotte}, quantum chaos~\cite{Jalabert}, and the
physics of dephasing~\cite{Zanardi, Fazio2}. Using this connection
and a combination of field theoretic tools, we compute exactly and
analytically $G(t)$ for global and local quantum quenches. In both
cases, we characterize the fluctuations of the work and  its
probability distribution. Interestingly, we show that for a local
quench starting at the quantum critical point the function $P(W)$
displays an edge singularity.

Let us start by briefly discussing the relation between the
characteristic function $G(t)$ and the Loschmidt echo. For a generic
quench $H(g_0) \rightarrow H(g_1)$, the Loschmidt echo is defined as
${\cal L}(t)=\mid {\cal G}(t) \mid^2$, where the amplitude ${\cal
G}$ is given by
\begin{eqnarray}
{\cal G}(t)=\left\langle e^{iH(g_0)t}e^{-iH(g_1)t} \right\rangle.
\end{eqnarray}
Here $H(g_0)$ and $H(g_1)$ are the initial and final Hamiltonian
respectively, and the average is taken with respect to the initial
equilibrium density matrix $\rho_0=\exp[-\beta H(g_0)]/Z$. The
Loschmidt echo can be seen as a  measure of the sensitivity of the
system to the quench.
The connection with $P(W)$ emerges by
noticing that for a generic quench the characterization of the work
done on the system requires two energy measurements: one before and
one after it~\cite{Kurchan,Talkner}. If the results of such
measurements are $\tilde{E}$ and $E$, the work done is then
$W=E-\tilde{E}$. Hence if $\mid \Phi_n \rangle$ are the eigenstates
of energy $E_n$ of $H(g_1)$, we have that
\begin{eqnarray}
P(W)=\sum_{n,m} \delta(W-(E_n-\tilde{E}_m)) \mid \langle \Phi_n \mid
\Psi_m \rangle \mid^2 P_m,
\end{eqnarray}
where $\mid \Psi_m \rangle$ are the eigenstates of $H_0$ with energy
$\tilde{E}_m$, and $P_m=\exp[-\beta \tilde{E}_m]/Z$. The
characteristic function is then $G(t)=\sum_n
e^{i(E_n-\tilde{E}_m)t}\mid ~\langle \Phi_n \mid \Psi_m \rangle
\mid^2 P_m$, which is readily recognized to be the complex conjugate
of the amplitude defining the Loschmidt echo $G(t)=[{\cal G}(t)]^*$.
This equality is actually a special case of the generalized quantum
Jarzynski equality~\cite{Kurchan,Jarzynski} recently derived in
Ref.~\onlinecite{Talkner} for problems in which $g$ is taken from
$g_0$ to $g_1$ in a finite time interval along a  generic path
$g(t)$.
The Loschmidt echo ${\cal L}(t)$ can be in principle measured by
studying the dephasing of an auxiliary two level system coupled to
the system of interest~\cite{Jalabert,Zanardi,Fazio2}. In the same
setup, the probability distribution $P(W)$ can be directly extracted
from the absorbtion spectra associated to optical transitions in the
auxiliary two level system~\cite{Silva}.

For a global quench the work done is extensive. Therefore in the
thermodynamic limit the probability distribution $P(w)$ of the work
per unit volume $w=W/V$ will be a strongly peaked function, with
fluctuations scaling as $1/\sqrt{V}$. This suggests that $P$  is a
nontrivial function only for small systems or for local quenches.
Despite this fact, it is interesting and instructive to study the
work statistics for a paradigmatic example: a zero temperature
global quench of the transverse field in a quantum Ising
chain~\cite{Sachdev}. The latter is defined by the Hamiltonian
\begin{eqnarray}
H_0=-J\sum_i \sigma^{x}_i \sigma^{x}_{i+1}+g
\sigma_i^z,\label{Ising}
\end{eqnarray}
where $\sigma^{x,z}_i$ are the spin operators at site $i$, $J$ is an
overall energy scale (below we set $J=1$) and $g$ is the strength of
the transverse field. The one dimensional quantum Ising model is the
prototypical, exactly solvable example of a quantum phase
transition, with a quantum critical point at $g_c=1$ separating two
mutually dual gapped phases, a quantum paramagnetic one ($g>g_c$)
and a ferromagnetic one ($g<g_c$).

Let us now consider a global change at time $t=0$ of the transverse
field from an initial value $g_0$ to a final one $g_1$. The analysis
of the Loschmidt echo can be efficiently performed after a
Jordan-Wigner transformation~\cite{Sachdev}. In the fermionic
representation, the Hamiltonian Eq.(\ref{Ising}) takes the simple
form
\begin{eqnarray}
H(g)&=&\sum_{k>0} (g-\cos(k))(c^{\dagger}_k
c_k-c_{-k}c^{\dagger}_{-k})\nonumber \\ &+&i\sin(k) (c^{\dagger}_k
c^{\dagger}_{-k} - c_{-k} c_{k}),
\end{eqnarray}
where $c_k$ are fermionic operators. The diagonal form $H=\sum_{k>0}
E_k (\gamma^{\dagger}_k \gamma_k-\gamma_{-k}
\gamma^{\dagger}_{-k})$, with energies
$E_k=\sqrt{(g-\cos(k))^2+\sin(k)^2}$, is achieved after a Bogoliubov
rotation $c_{k}=u_k(g) \gamma_k-iv_k(g) \gamma^{\dagger}_{-k}$,
$c^{\dagger}_{-k}=u_k(g)\gamma^{\dagger}_{-k}-iv_k(g) \gamma_k$. The
coefficients are given by
\begin{eqnarray}
u_k(g)=\cos(\theta_k(g))\;\;\;\;\;\;v_k(g)=\sin(\theta_k(g)),
\end{eqnarray}
where $\tan(2\theta_k(g))=\sin(k)/(g-\cos(k))$. In this
representation, the Loschmidt echo for the quantum Ising model
following both a global and a local quench of $g$ has been recently
shown to the expressible in terms of matrix determinants, which were
afterwards analyzed numerically~\cite{Zanardi,Fazio2}. Below, we
compute analytically the Loschmidt echo employing field theoretic
tools, which, in contrast to previous approaches, have the advantage
of giving a clear insight on the physics of the problem.

Our first task is to express the ground state $\mid \Psi_0 \rangle$
of energy $E_0$ of the initial Hamiltonian $H(g_0)$ in terms of the
eigenmodes $\gamma_k$ diagonalizing $H(g_1)$. If we call $\eta_k$
the eigenmodes of $H_0$ it is easy to see that
$\eta_k=U_k\gamma_k-iV_k \gamma^{\dagger}_{-k}$, where
\begin{eqnarray}
U_k=u_k(g_0)u_k(g_1)+v_k(g_0)v_k(g_1), \\
V_k=u_k(g_0)v_k(g_1)-v_k(g_0)u_k(g_1).
\end{eqnarray}
Hence the equation $\eta_k \mid \Psi_0 \rangle=0$, characterizing
our initial state can be easily solved giving
\begin{eqnarray}\label{boundary}
\mid \Psi_0 \rangle=\frac{1}{{\cal N}} \exp\left[ i
\sum_{k>0}\frac{V_k}{U_k} \gamma^{\dagger}_{k} \gamma^{\dagger}_{-k}
\right] \mid 0 \rangle,
\end{eqnarray}
where ${\cal N}$ is the normalization constant, and $\mid 0 \rangle$
is the vacuum of the fermions $\gamma_k$. The structure of this
state closely resembles that of integrable boundary states
encountered in statistical field theory. In particular, the
amplitude ${\cal G}$ is given by
\begin{eqnarray}~\label{result}
&&{\cal G}(t)= e^{i E_0 t}\langle \Psi_0 \mid e^{-i H(g_1) t} \mid
\Psi_0 \rangle  \\
&&= \frac{e^{-i \delta E t}}{{\cal N}^2} \langle 0 \mid e^{\sum_k
B^*(k) \gamma_k \gamma_{-k}} e^{\sum_k B(k)e^{-2iE_k t}
\gamma_{-k}^{\dagger}\gamma_{k}^{\dagger}} \mid 0 \ \rangle
\nonumber,
\end{eqnarray}
where $B(k)=-iV_k/U_k$, and $\delta E=E_1-E_0$, where $E_1$ is the
ground state energy of $H(g_1)$. Up to an irrelevant prefactor,
Eq.(\ref{result}) maps after a Wick rotation $it \rightarrow \tau$
onto the partition function of a two dimensional classical Ising
model constrained on a cylinder of height $\tau$ and with boundary
conditions on the two ends described by $\mid \Psi_0 \rangle$.
Hence, using techniques originally developed for integrable boundary
states~\cite{Mussardo} it is easy to obtain
\begin{eqnarray}~\label{result1}
{\cal G}(t)=e^{-i\delta E t} e^{L\int_0^{\pi} \frac{dk}{2\pi}
\log\left[\frac{1+ \mid B(k) \mid^2e^{-2iE_k t}}{ 1+ \mid B(k)
\mid^2}\right]},
\end{eqnarray}
where $L$ is the linear size of the chain.

We can now derive all the cumulants $C_n$ of the probability
distribution of the work per unit length $P(w)$ by expanding in
power series $\log(G(t/L))=\sum_{n=0}^{+\infty} C_n (it)^n/n!$. As
expected, from Eq.(\ref{result1}) we obtain that as $L$ grows, $C_n
\propto 1/L^{n-1}$, leading to the suppression of all fluctuations
in the thermodynamic limit. In order to study the effects associated
to the presence of the quantum critical point, let us focus on the
dependence on $g_0$ and $g_1$ of the average work per unit length
and of its variance. The first is in general given by $\langle w
\rangle=\delta E/L +w_e$, where the excess work is $w_e~\geq~0$, in
agreement with standard thermodynamic relations. The latter and the
variance have a particularly simple form if the final transverse
field is large $g_1 \gg 1$, in which case
$w_e = g_1 (1+g_0^2-\mid g_0^2-1\mid)/4g_0^2$
and $ \langle (\Delta w)^2 \rangle = (g_1^2)/L (g_0^4+4g_0^2-3-{\rm
sign}[g_0^2-1](g_0^4-4g_0^2+3))/8g_0^4$.
These functions are plotted in Fig.~\ref{Fig1a}, where one may see
that both $w_e$ and $\langle (\Delta w)^2 \rangle$ signal the
presence of the quantum critical point with discontinuities in their
derivative at $g=1$. More striking universal effects associated to
the quantum critical point are observed by studying  the asymptotics
of ${\cal G}(t)$ at long times, governed by the long wavelength
modes. If for example one looks at $g_0, g_1 \neq 1$ one may easily
obtain
\begin{eqnarray}
{\cal G}(t) \simeq {\cal G}^{\infty} \exp\left[{\cal A}\;\; m_1 L
\left(\frac{m_1-m_0}{m_1} \right)^2 \frac{e^{-2im_1 t}}{(i m
t)^{\frac{3}{2}}} \right],
\end{eqnarray}
where ${\cal G}^{\infty}$ is the asymptotic value attained by ${\cal
G}$ (signaling a delta function peak in $P(w)$), ${\cal A}$ is a
constant and $m_i=\mid g_i - 1 \mid$. Passing again to imaginary
time, the dependence on $t$ in the exponential corresponds to the
dependence on the thickness of the free energy of the Ising model on
a cylinder, which away from criticality is exponentially cutoff by
the correlation length $\xi = 1/m_1$. At criticality, of course, it
becomes a power-law as a result of the establishment of long range
correlations. A more detailed study of the statistics of $P(W)$ for
global quenches will be reported elsewhere~\cite{Silva}.

{\begin{figure} 
\epsfxsize=6cm \centerline{\epsfbox{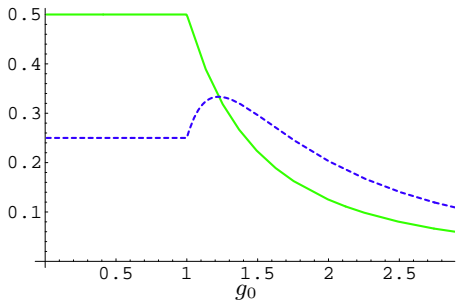}} \caption{A plot of
the average excess work per unit length $w_e/g_1$ (full line) and of
the variance $L/g_1^2 \langle (\Delta w)^2 \rangle$ (dashed line)
vs. $g_0$ for a global quench from $g_0$ to $g_1 \gg 1$. The
presence of the quantum critical point at $g_0=1$ is signaled by a
discontinuity in the derivative. } \label{Fig1a}
\end{figure}}


Let us now pass to a much more interesting situation in which we
expect the work done on the system to show nontrivial fluctuations:
a local quench of the Hamiltonian from $H_0=H(g)$ to $H_0+V$, where
\begin{eqnarray}
V=-\delta g \sigma^z_0.
\end{eqnarray}
In order to capture the main differences with the previous case, let
us start by considering the case $\delta g \ll 1$ and evaluate
${\cal G}(t)$ within a second order cumulant expansion
\begin{eqnarray}\label{cumulant}
{\cal G}(t)&=&\left\langle e^{iH_0t}\;e^{-i(H_0+V)t}
\right\rangle=\left\langle Te^{-i\int_0^t\;dt'\;V(t')}
\right\rangle\nonumber \\
&=& e^{-i \langle V \rangle t} e^{-\frac{1}{2} \int_0^t dt_1\;dt_2
\langle T[V(t_1)\;V(t_2)\;] \rangle},
\end{eqnarray}
where $V(t)=\exp[i H_0 t] V \exp[-i H_0 t]$. Using the fermionic
representation of the spin operators we obtain
\begin{eqnarray}
V=\frac{\delta g}{L} \sum_{k,k'} \left[ c_k c^{\dagger}_{k'}-
c^{\dagger}_k c_{k'} \right].
\end{eqnarray}
Hence writing $V$ in terms of the eigenmodes $\eta_k$ of $H_0$ and
substituting in Eq.(\ref{cumulant}), with the help of Wick theorem
we obtain
\begin{eqnarray}
{\cal G}(t)\simeq e^{-i\;\delta\!E\; t} e^{-f(t)},
\end{eqnarray}
Here the energy shift $\delta E$ is given by
\begin{eqnarray}
\delta E&=&  \delta g \int_{-\pi}^{\pi}
\frac{dk}{2\pi}\cos(2\theta_k(g))-\frac{(\delta g)^2}{2}
\int_{-\pi}^{\pi} \frac{dk dk'}{(2\pi)^2} \frac{V(k,k')}{E_k+E_k'}
\nonumber
\end{eqnarray}
where
$V(k,k')=\sin(2\theta_k)\sin(2\theta_{k'})+4\cos(\theta_k)^2\sin(\theta_k')^2$.

The most important information on the statistics of the work done on
the system is contained in
\begin{eqnarray}
f(t)\!=\!\frac{(\delta g)^2}{2} \!\!\int _{-\pi}^{\pi}\!\! \frac{dk
dk'}{(2\pi)^2}
\frac{V(k,k')}{(E_k+E_k')^2}\left(1-e^{-i(E_k+E_{k'})t} \right)
\end{eqnarray}
From this expression we may again estimate the various cumulants of
$P(W)$ by expanding in power series $f(t)$. In particular, the
variance is given close to the critical point by
\begin{eqnarray}
\langle (\Delta W)^2 \rangle \!\!=\!\!
%
\left(\frac{\delta g}{2\pi}\right)^2 \!\!\!(2(1+\pi^2) -
(g-1)(2+\log\left[\frac{\mid g-1 \mid}{8}\right]))\nonumber
\end{eqnarray}
Despite the fact that this function has a logarithmic singularity of
the first derivative at $g=1$, as originally found in studies of
dephasing~\cite{Fazio2}, it is important to notice that the integral
leading to this expression gets contributions from all frequencies
(not just small $k$). Hence universality does not emerge
substantially.

In order to obtain information on universal effects one has to study
the asymptotics of ${\cal G}(t)$ for long times. This can be done by
looking at the asymptotic value attained by $f$ at infinity
$f_{\infty}=(\delta g)^2/2 \int dk dk'/(2\pi)^2
V(k,k')/(E_k+E_{k'})^2$. Close to the quantum critical point $g
\simeq 1$, this is given by
\begin{eqnarray}
f_{\infty}\approx \left[\frac{\delta g}{2\pi} \right]^2 \log
\left[\frac{1}{\mid g-1 \mid}\right].
\end{eqnarray}
Hence as $t\rightarrow +\infty$ we have
\begin{eqnarray}
{\cal L}(t) \simeq \mid g-1 \mid^{2\left(\frac{\delta g}{2\pi}
\right)^2}.
\end{eqnarray}
The Loschmidt echo vanishes at the quantum critical point with a
cusp singularity. As shown below, the vanishing of the Loschmidt
echo is the result of an orthogonality catastrophe, originating from
the changing of a local scattering potential in a nontrivial, yet
gapless, effective fermionic system. In particular, if we set $g=1$
the long time decay of ${\cal G}$ is a power law
\begin{eqnarray}
{\cal G}(t) \approx e^{-i\delta E t} (it) ^{-\left(\frac{\delta
g}{2\pi} \right)^2}.
\end{eqnarray}
Hence we expect the probability distribution $P(W)$ to display an
edge singularity
\begin{eqnarray}
P(W) \approx \theta(W-\delta E) (W-\delta E)^{\left(\frac{\delta
g}{2\pi} \right)^2-1}.
\end{eqnarray}

This expectation is readily confirmed by the exact solution of the
problem for a local quench starting at the critical point $g=1$.
This can be obtained employing the scaling limit of the quantum
Ising model in the Majorana representation
\begin{eqnarray}
H_0[\varphi,\bar{\varphi}]=\int dr\;i\;
m\;\varphi\;\bar{\varphi}-\frac{i}{2}\varphi\;
\partial_r\; \varphi+\frac{i}{2}\bar{\varphi}\; \partial_r\;
\bar{\varphi},
\end{eqnarray}
where $\varphi$ and $\bar{\varphi}$ are Majorana fermions, and the
mass is related to the transverse field by $m=g-1$. The quench
consists in going from $H_0$ to $H_0+V$, where
$V[\varphi,\bar{\varphi}]=i\;\delta
m\;\varphi(0)\;\bar{\varphi}(0)$.

In order to compute ${\cal G}$ at criticality ($m=0$) let us use a
trick due to Itzykson and Zuber~\cite{Itzykson}. We start by
computing $\left[{\cal G}(t)\right]^2$. Introducing two copies of
the Majorana fermions, $\varphi_{1,2}$ and $\bar{\varphi}_{1,2}$, we
have
\begin{eqnarray}
[{\cal G}(t)]^2=\left\langle e^{i {\cal H}_0 t}e^{-i ({\cal
H}_0+{\cal V})t} \right\rangle,
\end{eqnarray}
where ${\cal
H}_0=H_0[\varphi_1,\bar{\varphi_1}]+H_0[\varphi_2,\bar{\varphi_2}]$
and ${\cal
V}=V[\varphi_1,\bar{\varphi_1}]+V[\varphi_2,\bar{\varphi_2}]$. The
most elegant way to proceed consists now in combining the Majorana
fermions into Dirac fermions
$\Psi_{R}=(\varphi_1+i\varphi_2)/\sqrt{2}$ and
$\Psi_{L}=(\bar{\varphi}_1+i\bar{\varphi}_2)/\sqrt{2}$, and then
introducing a pair of nonlocal operators~\cite{Kane} defined as
$\Psi_{+}(r)=(\Psi_{R}(r)+\Psi_L(-r))/\sqrt{2}$, and
$\Psi_{-}(r)=(\Psi_{R}(r)-\Psi_L(-r))/\sqrt{2}i$.
In terms of these
\begin{eqnarray}
{\cal H}_0 &=& \int dr \Psi^{\dagger}_+ (-i\partial_r) \Psi_+
+\Psi^{\dagger}_- (-i\partial_r) \Psi_- ,\\
{\cal V} &=& \delta m (\Psi^{\dagger}_+(0) \Psi_+(0)-
\Psi_-^{\dagger}(0)\Psi_-(0)).
\end{eqnarray}

Physically, it is now evident that we have two chiral modes subject
to local potential scattering of opposite sign characterized by
phase shifts $\pm \delta=\pm \delta m/2$. The computation of ${\cal
G}^2$ is now a standard problem solvable by
bosonization~\cite{Schotte,note}. We
find that
\begin{eqnarray}
{\cal G}(t) = \left[\frac{1}{1+i t }
\right]^{\left(\frac{\delta}{\pi}\right)^2}.
\end{eqnarray}
The complex conjugate of this expression is readily recognized to be
the characteristic function of the Gamma probability distribution
\begin{eqnarray}
P(w)=\frac{ w^{\left(\frac{\delta}{\pi}\right)^2-1}
\;e^{-w}}{\Gamma\left[\left(\frac{\delta}{\pi}\right)^2 \right]},
\end{eqnarray}
which indeed displays an edge singularity with a exponent
$(\delta/\pi)^2=(\delta g/(2\pi))^2$ consistent with the one
obtained by the cumulant expansion.

In conclusion, after elucidating the connection between the
probability distribution $P(W)$ of the work done on a system in a
quantum quench and the Loschmidt echo, we characterized $P(W)$ for
global and local quenches of the transverse field in a quantum Ising
chain. As mentioned before, the experimental measurement of $P(W)$
requires the realization of an optical absorbtion experiment in a
fully controllable setting. Recent proposals for the realization of
quantum spin chains using bosonic atoms in optical
lattices~\cite{Lukin} give a possible, concrete way to pursue this
goal with the available experimental tools.

I would like to thank G. Mussardo for important discussions, help
and encouragement throughout this project and R. Fazio for
stimulating my interest on the Loschmidt echo. I would also like to
thank  V. Kravtsov, A. Nersesyan, D.Rossini and G. Santoro for
discussions.


\begin{thebibliography}{10}
\bibitem{note} Here we set the ultraviolet cutoff
$\alpha \simeq J =1$.
\end{thebibliography}
\end{document}